\documentclass[pra, twocolumn,  notitlepage, showpacs, floatfix]{revtex4-1}
\usepackage{amsmath}
\usepackage{graphicx}
\usepackage{amsfonts}
\usepackage{amssymb}
\usepackage{mathtools}
\usepackage{epstopdf}
\usepackage{bbold}
\usepackage{color}

\usepackage{MnSymbol}

\DeclareMathAlphabet\mathbfcal{OMS}{cmsy}{b}{n}

\DeclareMathOperator{\Tr}{Tr}

\newcommand{\ket}[1]{| #1 \rangle}
\newcommand{\bra}[1]{\langle #1 |}
\newcommand{\braket}[2]{\langle #1 | #2 \rangle }

\newcommand{\tr}{\mathrm{Tr}}
\renewcommand{\t}[1]{\mathrm{#1}}

\begin{document}

\title{Discrete-to-continuous transition in quantum phase estimation}
\author{Wojciech Rz\c{a}dkowski}
\affiliation{Faculty of Physics, University of Warsaw,  ul. Pasteura 5, PL-02-093 Warszawa, Poland}
\author{Rafa\l{} Demkowicz-Dobrza\'nski}
\affiliation{Faculty of Physics, University of Warsaw,  ul. Pasteura 5, PL-02-093 Warszawa, Poland}

\begin{abstract}
We analyze the problem of quantum phase estimation where the set of allowed
phases forms a discrete $N$ element subset of the whole $[0,2\pi]$ interval,
$\varphi_n = 2\pi n/N$, $n=0,\dots N-1$ and study the discrete-to-continuous
transition $N\rightarrow\infty$ for various cost functions as well as the mutual information. 
We also analyze the relation between the problems of phase discrimination and 
estimation by considering
a step cost functions of a given width $\sigma$ around the true estimated value.
We show that in general a direct application of
the theory of covariant measurements for a discrete subgroup of the $U(1)$ group
leads to suboptimal strategies due to an implicit requirement of estimating only the phases
that appear in the prior distribution. We develop the theory of  sub-covariant measurements to remedy this situation and demonstrate truly optimal estimation strategies when performing transition from a discrete to the continuous phase estimation regime.
\end{abstract}
\maketitle

\section{Introduction}
Quantum phase estimation is a paradigmatic model capturing the essence of all interferometric experiments irrespectively whether they are performed using atoms or light \cite{Hariharan2003, Cronin2009}. This is at the same time the best studied model in the field of theoretical quantum estimation theory
 \cite{Helstrom1976, Holevo1982}
 and lies at the very foundations of the whole field of quantum metrology
  \cite{Giovannetti2011, Toth2014,  Demkowicz2015, Pezze2016}. It has been studied both in idealized scenarios \cite{Caves1981, Yurke1986, Holland1993, Dowling1998, Berry2000} as well as in presence of various decoherence effects \cite{Huelga1997, Shaji2007, Dorner2008, Knysh2010, Genoni2011, Escher2011, Demkowicz2012}.

The problem has also been analyzed using two different conceptual perspectives:
the frequentist approach and the Bayesian one. The first approach focuses on
scenarios where the experiment is repeated many times, and provides useful
bounds on the performance of optimal estimator in the form of  the famous
Cram{\'e}r-Rao bound.
The main tool here is the concept of Fisher information or its quantum
generalization---the Quantum Fisher Information (QFI) \cite{Helstrom1976,
Braunstein1994}. The second approach, requires specification of the prior
distribution
describing the knowledge on the parameter to be estimated but
is capable of providing operationally meaningful results dealing directly with single-shot experiments without the need to go into the limit of many independent experiment repetitions \cite{Berry2000, Demkowicz2011, Hall2012, Jarzyna2015}.

Within the quantum estimation theory, both approaches are applied to models where the phase parameter to be estimated is treated as continuous. In the frequentist approach, this is manifested
explicitly in the definition of the QFI, where derivatives with respect to the estimated parameter appear. In the Bayesian approach, one typically chooses a natural flat prior distribution for the phase
$p(\varphi)=1/2\pi$, representing our complete initial ignorance on the actual value of the phase.

In this paper we  analyze the phase estimation situation in case the set of allowed  phases
is discrete: $\varphi \in \mathcal{S}_N$, $\mathcal{S}_N = \{ n \theta, \theta =
2\pi/N, n=0,\dots, N-1 \}$, and analyze the transition to the continuous limit
$N \rightarrow \infty$.
Discretization of phase is often encountered in optical communication protocols,
where discrete set of phases is encoded in states of light in the so-called
phase-shift keying protocols \cite{Kitayama2014, Muller2015, Jarzyna2016}. In
metrological scenarios, such situations where the phase to be estimated
is ``quantized'' are not commonly encountered, but can be of relevance for
examples in models
where atoms passing through an optical cavity acquire a phase proportional to
the number of photons inside \cite{Brune2008}. This said, we admit that our main
motivation here is to understand the discrete-to-continuous transition from a
purely conceptual perspective as we find this important aspect of quantum phase
estimation surprisingly unexplored.

Clearly, the formulation of the problem, sets us immediately into the Bayesian framework, as the problem can be phrased in a Bayesian language by stating that the prior probability for phase estimation is simply $p(\varphi) = \frac{1}{N} \sum_{n=0}^{N-1} \delta(\varphi - n \theta)$.
One can argue that assuming a discrete set of allowed phases moves us from the
problem of estimation of phase to the problem of phase discrimination
\cite{Barnett2009}. Indeed, this can be viewed this way, but an important
element to be specified here is the explicit form of the cost function that we
assume in the problem.
If we choose a simple delta cost function penalizing us equally strongly
whenever we guess the wrong phase, we will indeed reduce our problem to the one
studied in the quantum discrimination literature \cite{Barnett2009}, and in a
sense the discrete to continuous transition for such models will be trivial as
discussed explicitly in Sec.~\ref{sec:discrimination}. Still, for any other cost
function, the transition will be nontrivial and one cannot directly utilize
known results from quantum state discrimination theory.

This paper is organized as follows. In Sec.~\ref{sec:discrete} we formulate the
problem of discrete phase estimation considered throughout this work. 
Secs.~\ref{sec:standard} and~\ref{sec:discrimination} provide details of 
discrete-to-continuous transition while focusing on two popular cost functions:
the $4 \sin^2\frac{\varphi-\tilde{\varphi}}{2}$ cost function commonly used in 
continuous phase estimation problems and
the fixed interval cost function more natural in discrimination-like problems, 
respectively. After discussing these two examples, in Sec.~\ref{sec:general} we
discuss behavior of the optimal estimation protocols without assuming any 
particular choice of a cost function and we indicate
conditions under which going beyond standard covariant measurements may lead to 
a reduced estimation cost. 
Sec.~\ref{sec:subcovariant} provides a more abstract and formal consideration  
of the use of sub-covariant measurements
in a general problem of sub-group element estimation, of which discrete phase estimation is a special case.  In
Sec.~\ref{sec:algorithm} we describe a numerical framework
within which one can numerically optimize a general sub-group estimation problem using the idea of sub-covariant measurements.
The discrete-to-continuous transition is studied using a different figure of 
merit---mutual information in Sec.~\ref{sec:mutual}.
The last Sec.~\ref{sec:conclusion} concludes the paper.

\section{Discrete phase estimation}
\label{sec:discrete}
Let us consider a $D$-dimensional quantum system,  where the phase parameter
$\varphi$ is encoded
on a general input probe state $\ket{\varphi_0} = \sum_{k=0}^{D-1} c_k \ket{k}$ as follows:
\begin{equation}
\ket{\varphi} = U_\varphi \ket{\varphi_0} = \sum_{k=0}^{D-1} c_k e^{i k \varphi} \ket{k}.
\end{equation}
This model may represent the evolution of a single mode state of light
traveling through a phase delaying element, in which case $\ket{k}$ represents a
$k$-photon state.
Equivalently, in a more physically meaningful scenario, one can think of a two
arm interferometer where a ($D-1$)-photon state has been split into two arms of
the interferometer and $\varphi$ represent the relative phase delay for the
light traveling through the upper and lower arms. In this case
$\ket{k}$ should be understood  as $\ket{k} \otimes \ket{D-1-k}$, representing
the state where $k$ photons go through the upper and $D-1-k$ photons go through
the lower arm of the interferometer. Analogously in atomic Ramsey
interferometry, we would think of $\ket{k}$ as representing the state with $k$
atoms in the exited and  $D-1-k$ atoms in the ground state \cite{Ramsey1980}.

Let $C(\varphi, \tilde{\varphi})$ be the cost function representing the cost of estimating the
value $\tilde{\varphi}$ while the true value of the phase is $\varphi$.
 In what follows, we will naturally assume $C$ to depend only on the difference of the phases $
 C(\varphi, \tilde{\varphi}) = C(\varphi-  \tilde{\varphi})$.
Since in discrete phase estimation, allowed phases are  $\varphi_n = n \theta$, 
the average cost 
is given by:
\begin{equation}
\bar{C} = \frac{1}{N} \sum_{n=0}^{N-1}
\sumint\limits_m
\tr\left(\ket{\varphi_n} \bra{\varphi_n} \Pi_m \right)C(\varphi_n - \tilde{\varphi}(m)),
\end{equation}
where $\{\Pi_m\}$ represent a generalized POVM measurement \cite{Nielsen2000},
$\Pi_m \geq 0$, $\sumint_m \Pi_m = \openone$,
while $\tilde{\varphi}$ is an estimator function assigning a given value of
phase  $\tilde{\varphi}(m)$ to a given measurement outcome $m$.
Determining the optimal discrete phase estimation protocol amounts to  minimizing the above quantity over
the input state $\ket{\varphi_0}$, measurement $\{\Pi_m\}$ and the estimator $\tilde{\varphi}$.

Even though the above optimization appears extremely challenging, the symmetry
of the problem
helps to simplify it considerably.
In case of continuous phase estimation, the problem has a natural symmetry with respect to
phase shifts, or more formally $U(1)$ group. The prior distribution $p(\varphi) = 1/2\pi$ is
invariant under phase shifts, as well as the cost function $C(\tilde{\varphi}+\xi, \varphi+\xi)= C(\tilde{\varphi}, \varphi)$, while the family
 of states $\ket{\varphi}$
is obtained by acting with a unitary representation $U_\varphi$ of the group on the probe state
$\ket{\varphi_0}$. This is an example of a covariant estimation problem \cite{Holevo1982}.
In this case, it has been proven that while looking for the optimal estimation scenarios
one can restrict oneself to the so-called covariant measurements, where
respective measurement operators are generated from a single seed POVM by the action of the representation $U_\varphi$: $\Pi_{\tilde{\varphi}} = U_{\tilde{\varphi}} \Pi_0 U_{\tilde{\varphi}}^\dagger$.
Note that measurement operators are labeled by a continuous parameter
$\tilde{\varphi}$, where it is implicitly assumed that the index represents also
the value of the estimated phase, given a particular measurement result occurs.
Thanks to the use of covariant measurements the formula for the average cost
function simplifies to:
\begin{equation}
\label{eq:costcon}
\bar{C} = \frac{1}{2\pi}\int \t{d} \varphi \tr(\ket{\varphi} \bra{\varphi} \Pi_0) C(\varphi).
\end{equation}
One of standard choices for the cost function in phase estimation literature is
$C(\varphi) = 4 \sin^2 \frac{\varphi}{2}$. This function is the simplest (in the sense of expansion into
a Fourier series in $\cos (m\varphi)$) non-trivial function that approximates
squared error for small phase deviations.  For the problem of phase estimation
with the above chosen cost function, minimization of the above formula over
$\ket{\varphi_0}$ and $\Pi_0$ can be done analytically \cite{Berry2000}, and
results in
$\Pi_0 = D\ket{e}\bra{e}$, $\ket{e} = \frac{1}{\sqrt{D}}\sum_{k=0}^{D-1} \ket{k}$ and
\begin{equation}
\ket{\varphi_0} =\sqrt{\frac{2}{D+1}} \sum_{k=0}^{D-1}
\sin\left(\frac{k+1}{D+1}\pi  \right)\ket{k},
\end{equation}
yielding the minimal cost $\bar{C} = 2\left[1-\cos\left(\frac{\pi}{D+1} \right) \right]$.

It would seem that when considering a discrete phase estimation problem,
 all one has to do is replace the whole $U(1)$ group with its discrete $N$-element subgroup $U_N$
 and proceed as before. Indeed the problem is covariant with respect to the $U_N$ group, so the discrete variant of \eqref{eq:costcon} reads
\begin{equation}
\label{eq:costdiscrete}
\bar{C} = \frac{1}{N} \sum_{n=0}^{N-1} \tr\left( \ket{\varphi_n}\bra{\varphi_n}\Pi_0 \right) C( \varphi_n).
\end{equation}
Since the above formula can be rewritten as
\begin{equation}
\label{eq:optstate}
\bar{C} = \bra{\varphi_0} A \ket{\varphi_0}, \quad A =
\frac{1}{N} \sum_{n=0}^{N-1} U_{\varphi_n} \Pi_{0} U_{\varphi_n}^\dagger
C(\varphi_n )
\end{equation}
the optimal input probe state can be determined as the eigenvector corresponding to the minimal eigenvalue of the matrix $A$,
while the optimal measurement can always be chosen as $\Pi_0 = D \ket{e} \bra{e}$ \cite{Holevo1982}.

As we discuss explicitly further on in the paper, this approach for  determining the optimal discrimination strategy
may sometimes lead to a counter-intuitive result that cost of discrete estimation for some finite $N$ appears to be larger
than that of the continuous case.
This apparent paradox may be understood if we realize an implicit assumption we
have
additionally made while moving from the continuous to a discrete case, while keeping the structure of covariant measurements.
Namely, we have  assumed that
estimated values $\tilde{\varphi}(m)$ are restricted to belong to the same set
$\mathcal{S}_N$ of phases that are being encoded on the input state. Depending
on the choice of the cost function, it might be the case that in order to
minimize $\bar{C}$ it is advantageous to estimate phases
outside $\mathcal{S}_N$. In what follows we will refer to such strategies as sub-covariant measurements.


\section{Standard phase estimation cost function}
\label{sec:standard}
We start by considering the cost function $C(\varphi)= 4 \sin^2 (\varphi/2)
= 2 (1- \cos\varphi)$ commonly encountered in continuous phase estimation problems.
Let us first restrict to proper covariant measurements, so assume $\tilde{\varphi} \in \mathcal{S}_N$,
and $\Pi_{\tilde{\varphi}} = U_{\tilde{\varphi}} \Pi_0 U_{\tilde{\varphi}}^\dagger$.
When substituting the explicit parametrization of the input state $\ket{\varphi_0} = \sum_{k}c_k \ket{k}$
to \eqref{eq:costdiscrete} the average cost reads:
\begin{equation}
\label{eq:costexplicit}
\bar{C} = 2- \frac{1}{N}\sum_{n=0}^{N-1}2 \cos (n \theta) \sum_{j,k=0}^{D-1} 
\left(\Pi_0\right)_{jk} c_jc_k^* e^{i(j-k)n\theta}.
\end{equation}
In case $N=D$ a simple choice
$\Pi_0 = D \ket{e}\bra{e}$, $\ket{\varphi_0} = \ket{e}$
leads to $\bar{C}=0$, as $\sum_{j=0}^{D-1} e^{i j n \theta} = \sum_{j=0}^{N-1}
e^{2 \pi i j n/N}= N \delta_{n,0}$---the discrete Fourier transform property.
Similarly $\bar{C}$ can easily be made $0$ for $N < D$. One can simply choose
the same measurement operator but
consider an input state supported on a $N$ dimensional subspace $\ket{\varphi_0} = \frac{1}{\sqrt{N}}\sum_{k=0}^{N-1} \ket{k}$ and the same argument follows. The above observation is in fact general. Irrespectively of the cost function for $N\leq D$ the cost can be made $0$. This is simply a manifestation of the fact
that the phase transformation is capable of generating up to $D$ orthogonal states which can be perfectly discriminated.

The first non-trivial case is therefore $N=D+1$. Due to $\cos (n \theta)$ factor while performing the sum over $n$ the following sums will appear
$\sum_{n=0}^{N-1} e^{i [(j-k) \pm 1] n 2\pi/N}$. This sums will be nonzero iff $(j-k) \pm 1 = 0$ (note that since $D=N-1$ they $|(j-k) \pm 1| < N$).
Hence only first off-diagonal terms will contribute to the cost. Now, observe that had we considered continuous case $N \rightarrow \infty$, the conclusions
would remain the same. This implies that the cost, measurement as well as the optimal state
will be the same as in the continuous case. This proves a rather surprising fact, that in this model estimating $D+1$ phases is equally
challenging as estimating a phase continuous parameter.  This also implies
that for $N \leq D$ and $N \geq D+1$ standard covariant measurement are optimal 
as they provide perfect discrimination for $N\leq D$ and
are known to be optimal in the continuous limit. Since all the sums we perform result in identical formulas as the ones in the
continuous case, then if indeed a class of more general measurements allowed to
further lower the cost then the same measurement could be used in the
continuous case and would lead to a contradiction with the known optimal result
for continuous phase estimation.

Let us also note that in order to solve this problem one could have also
resorted to methods of \cite{Demkowicz2011} where the problem of phase
estimation with arbitrary prior distribution has been analyzed for the case of
$4 \sin^2 \varphi/2$ cost function---in particular a discrete prior
distribution.

\section{Discrimination-like cost function}
\label{sec:discrimination}
Let us now consider the following step cost function: $C(\varphi) = 0$ for
$|\varphi|<\sigma/2$ and $1$ otherwise.
If we take the limit $\sigma \rightarrow 0$ then we face phase discrimination problem, where unless we guess the correct phase we
are penalized equally. From quantum state discrimination theory
\cite{Barnett2009} it is known that for $N$ arbitrary quantum states
$\ket{\psi_n}$ generated by any unitary
$\ket{\psi_n} = U^{n} \ket{\psi_0}$ such that $U^N = \openone$ the
optimal measurement minimizing the discrimination cost are the so-called
 square root measurements defined as follows: $\Pi_n = \frac{1}{N} \rho^{-1/2}
\ket{\psi_n} \bra{\psi_n} \rho^{-1/2}$, where
 $\rho = \frac{1}{N} \sum_{n=0}^{N-1}\ket{\psi_n}\bra{\psi_n}$.

In our case $U=U_\theta = \t{diag}(1, e^{i\theta},\dots,e^{i (D-1)\theta})$,
$\ket{\psi_n}=\ket{\varphi_n}$ and for $N \geq D$ we
obtain that for input state with positive coefficients $c_k > 0$, the optimal  square root measurement is just a standard covariant measurement $\Pi_{n\theta} = U_{n \theta} \Pi_0 U_{n \theta}^\dagger$
with $\Pi_0 = D \ket{e}\bra{e}$ (for other states we just need to correct for
the phases of coefficients $c_k$).
The optimal state is the one that minimizes the mean cost, which in
this case amounts to:
\begin{equation}
\bar{C} = 1 -  \frac{D}{N}|\braket{\varphi_0}{e}|^2 = 1- \frac{D}{N} \quad
\t{for}\ N \geq D,
\end{equation}
and the above optimal value is obtained for $\ket{\varphi_0} = \ket{e}$. For $N \leq D$ as before the cost is zero.

The above formula is valid only in the limit $\sigma \rightarrow 0$. For any finite $\sigma$ we expect the above formula to hold at least in the regime
where width of the cost function $\sigma$ is smaller than the separation between the phases  $\theta$, as in this case
the width of the cost function should not play a role.
At the point were more than one phase can fit into the width of the cost function the situation ceases to correspond to a discrimination problem
and the more phases fit into the cost function width the more ``estimation-like'' the problem becomes.

\begin{figure}[t]
\includegraphics[width=\columnwidth]{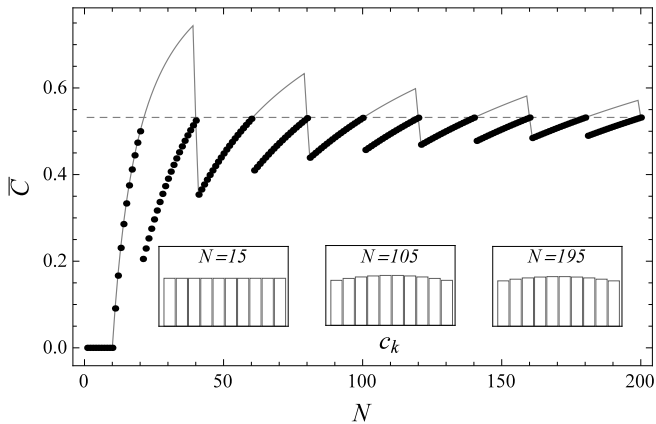}
\caption{Average cost $\bar{C}$ as a function of level of phase discretization (number of phases $N$) for
a $D=10$ dimensional system and the width of the step cost function $\sigma=\pi/10$.
Gray line corresponds to the minimal cost obtainable with standard covariant measurements.
Black dots correspond to the optimal strategy. In regimes where the optimal strategy outperforms the covariant one, the use of sub-covariant measurements is necessary. In this case improvement is due  to the possibility of estimating phases that lie in the middle between the phases actually encoded in the state. Dashed line represent the value of the cost for the continuous limit $ N \rightarrow \infty$.
The inset depicts coefficients $c_k$ of the optimal input state at a given $N$.}
\label{fig:discrimination}
\end{figure}
We first assume we restrict ourselves to covariant measurements. We take $\Pi_0 = D \ket{e}\bra{e}$ while
the optimal state we determine performing determining the vector corresponding to the minimal eigenvalue of  matrix $A$ as defined in
Eq.~~\eqref{eq:optstate}.
The results of the above procedure are depicted as a gray line in Fig.~\ref{fig:discrimination}, for an exemplary case
$D=10$ and $\sigma=\pi/10$. We can observe an apparently paradoxical behavior of 
the cost for some $N$ is above the
one achieved in the continuous limit, which could be phrased in a way that discriminating a subset of states is more difficult than the
whole set, a clear contradiction.

In order to remedy for this apparent paradox one  should go beyond the standard covariant measurement class and consider the possibility of estimating phases which are not inside the set of encoded phases. In the considered case it is enough to include shifted-covariant measurements where the estimated phases  correspond not to
$\varphi_n = n\theta$ but to values exactly in between the actual phases encoded in the state., i.e. $\varphi_n + \theta/2$.
In other words we should replace $\Pi_0$ with
$\Pi_{\frac{1}{2}\theta} = D U_{\frac{1}{2}\theta} \ket{e}\bra{e} U^\dagger_{\frac{1}{2}\theta}$. This is a special case of the
class of measurements we refer to as sub-covariant measurements, which are 
described in a formal way in Sec.~\ref{sec:subcovariant}.
Depending on the value of $N$ one should switch between the two strategies.
The resulting minimal cost is depicted as black dots in Fig.~\ref{fig:discrimination}.

The results are easy to understand.
For $N\leq 10$, $\bar{C}=0$ as expected. Then for $11\leq  N \leq 20$ the curve follows
the $1-\frac{D}{N}$ formula corresponding to the optimal phase discrimination problem.
At $N=21$, we observe a change due to the fact that the separation between the phases
$2\pi/21$ is smaller than $\sigma$ and hence two phases can fit into the width of the cost function.
The optimal measurement in this case corresponds to sub-covariant measurement.  Another transition happens
at $N=41$. This a point where $\sigma > 2 \theta$ and hence three phases can fit into one width of the cost function. At this point we recover the optimality of the standard covariant measurement.
This situation repeats itself at every $N = k \times 20 +1$, switching back and forth between the optimal covariant and shifted-covariant strategies.

In the continuous limit $N\rightarrow \infty$ we may of course use the standard covariant measurements. Plugging in $\Pi_0 = D \ket{e}\bra{e}$ into \eqref{eq:costcon} and integrating over $\varphi$ for our step function cost we get:
\begin{equation}
\bar{C} = \bra{\varphi_0} B \ket{\varphi_0}, \quad B_{jk} = \begin{cases}
1- \frac{\sigma}{2\pi} & j =k \\
-\frac{1}{\pi(j-k)} \sin[\sigma(j-k)/2] & j\neq k.
\end{cases}
\end{equation}
The eigenproblem of matrix $B$ is well known from Fourier
analysis~\cite{Slepian1978} and has also been studied in the context of quantum
communication~\cite{Hall1991,Hall1993}. The optimal input state is the
eigenvector of matrix $B$ corresponding to the smallest eigenvalue which is then
the minimal cost. The coefficients of this optimal eigenvector
form a Discrete Prolate Spheroidal
Sequence (DPSS)~\cite{Slepian1978}. The smallest eigenvalue (minimal cost)
corresponding to the DPSS eigenvector is plotted as a dashed line in
Fig.~\ref{fig:discrimination}.
Note that thanks to the use of sub-covariant measurement the optimal cost for
finite $N$ never exceeds the one corresponding to the continuous limit which
should be the case as the continuous phase estimation should be never be less
difficult than the discrete one.

\section{General cost function}
\label{sec:general}
Having studied this particular two examples, let us move on to a more general discussion
 where we will  provide some general statements without specifying a particular form of the cost function.
 We consider a general cost function given by
\begin{equation}
\label{eq:generalcost}
 C(\varphi)=\sum\limits_{m=0}^M
 \alpha_m \cos(m\varphi),
\end{equation}
where the Fourier series structure of the parametrization reflects the
periodicity of the function and we will refer to $M$ as order of the cost
function. For the cost function to be meaningful, the coefficients $\alpha_m$ 
should be such to make $C(\varphi) \geq 0$, and $C(0)=0$.
Let us first restrict to proper covariant measurements, so assume
$\tilde{\varphi} \in \mathcal{S}_N$,
and $\Pi_{\tilde{\varphi}} = U_{\tilde{\varphi}} \Pi_0
U_{\tilde{\varphi}}^\dagger$. After inserting the cost
function~(\ref{eq:generalcost}) and the explicit parametrization of the input
state $\ket{\varphi_0} = \sum_{k}c_k \ket{k}$
to \eqref{eq:costdiscrete} the average cost reads:
\begin{equation}
\label{eq:costexplicitgeneral}
\bar{C} = \sum\limits_{m=0}^M\alpha_m\frac{1}{N}\sum_{n=0}^{N-1} \cos ( m n
\theta) \sum_{j,k=0}^{D-1} \left(\Pi_0\right)_{jk} c_jc_k^* e^{i(j-k)n\theta}.
\end{equation}
\begin{figure}[t!]
\includegraphics[width=\columnwidth]{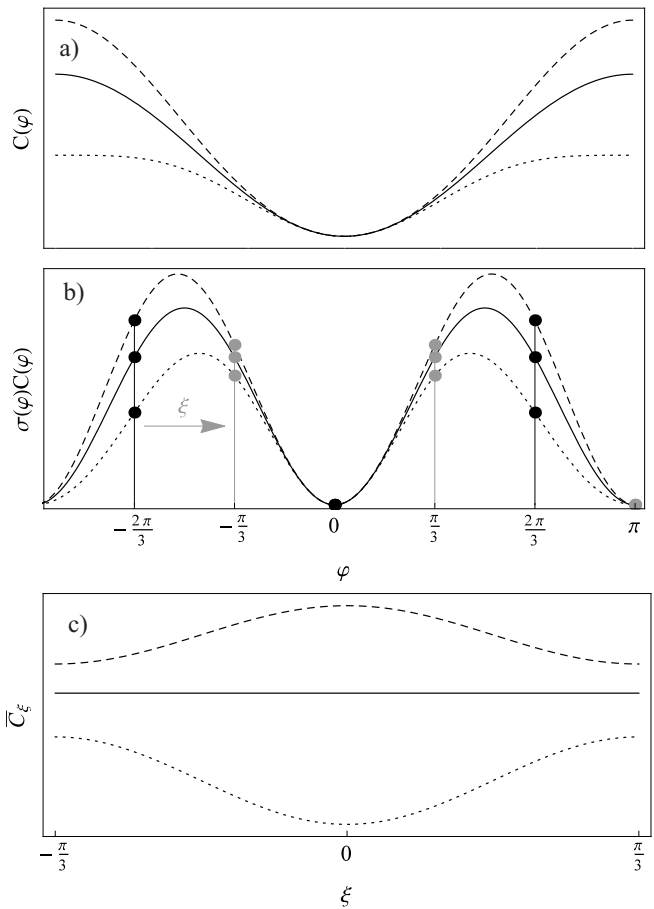}
\caption{Difference in performance between covariant and shifted-covariant measurements
in discrete phase estimation illustrated for a simple case of a qubit $D=2$ and three phases $N=3$.
Generic behavior is depicted using three representative cost functions plotted 
in a):
the standard cost function $C_0 = 4\sin^2\frac{\varphi}{2}$ (solid) as well as exemplary second order cost functions which approximate $\varphi^2$ for small $\varphi$ but are above or below the standard cost function respectively: $C_1= \frac{15}{6} - \frac{8}{3} \cos{\varphi} + \frac{1}{6} \cos{2 \varphi}$ (dashed), $C_2 = \frac{5}{4} - \cos\varphi - \frac{1}{4} \cos(2 \varphi)$ (dotted).
In b) difference between  covariant measurement and shifted-covariant measurement, where the shift parameter is chosen to be $\xi=\pi/3$, is depicted as different sampling of the probability weighted cost function $\sigma(\varphi) C(\varphi)$.
In general, for symmetric cost and probability distributions, the average cost $\bar{C}_\xi$ is going to have two extremal
points at $\xi=0$ (covariant measurement) and $\xi = \pi/N = \pi/3$ (shifted-covariant measurement).
Depending on the character of the average cost function the minimum is achieved at one point or the other.
In the presented example, the shift is advantageous for $C_1$ cost function but 
not for $C_2$ and is irrelevant for $C_0$ since for this first order cost 
function $D=2$ and $N=3$ imply that the continuous phase estimation limit has 
already been reached.
}
\label{fig:probing}
\end{figure}

As already discussed when analyzing the standard phase estimation cost function,
in case $N \leq D$, the cost can always be made zero.

Let us now consider $N \geq D+1$. Due to $\cos ( m n \theta)$
factor
while performing the sum over $n$, the following sums will appear
$\sum_{n=0}^{N-1} e^{i [(j-k) \pm m ] n 2\pi/N}$. This sums will be nonzero iff
$(j-k) \pm m = 0, \pm N, \pm 2N, \dots$ 
If we now increase number of phases such that $N = D + M $, then
$|(j - k) \pm m| < N$ and hence only terms where $(j-k) \pm m = 0$, i.e. first $M$ off diagonal terms of the state,  will contribute.
This also means than increasing $N$ further will not change the resulting cost, as the same terms will enter in the formula.
%

Hence  we can draw a general conclusion -- after being given an arbitrary cost
function with order $M$, we predict the minimal cost to be zero for $N
\leq D$, and then increase up to $N = D +M$,
at which point it already reaches its continuous limit. This also 
implies
that for $N \leq D$ and $N \geq N+M$ standard covariant measurement will
suffice to reach the optimal performance.
However, one cannot a priori assume that covariant measurements will be also optimal in the regime $D < N < D + M$.

Let us now further analyze the structure of formula \eqref{eq:costexplicitgeneral}.
We can write it
in the following form:
\begin{equation}
\label{eq:withoutoffset}
 \bar C = \sum\limits_{n=0}^N \sigma(\varphi_n)C(\varphi_n),
\end{equation}
 where $\sigma(\varphi)=\tr\left(
\ket{\varphi}\bra{\varphi}\Pi_0 \right)/N$.
Intuitively, the continuous estimation corresponds to integrating the function
$\sigma(\varphi)C(\varphi)$ from $0$ to $2\pi$, while discrete estimation
corresponds to summing discretely probed values of this function. The simplest
step beyond the covariant strategy is to introduce a non-zero offset
$\xi$:
\begin{equation}
\label{eq:withoffset}
 \bar C_\xi = \sum\limits_{n=0}^N \sigma(\varphi_n+\xi)C(\varphi_n+\xi),
\end{equation}
i.e. to probe the function $\sigma(\varphi)C(\varphi)$ at different
values of $\varphi$. This physically is equivalent to providing estimated
phases outside the set of true phases, but still keeping an equal number of true
and estimated phases. This is what we referred to before as shifted-covariant 
measurement.

Due to natural symmetry of the problem considered, both $C(\varphi)$ and 
$\sigma(\varphi)$ are symmetric functions and hence also probability weighted 
cost function $\sigma(\varphi) C(\varphi)$.
This implies that $\bar{C}_\xi$ has an extremum at $\xi=0$. This also implies that another extremum should appear at $\xi = \pi/N$ corresponding to a shift by half the interval between the encoded phases---increasing or decreasing the value od $\xi$ around $\xi=\pi/N$ has the same effect on $\bar{C}_\xi$ as it can be seen as the same operation performed on original weighted cost function or on its reflection around $\varphi=0$ which is the same due to symmetry of the probability weighted cost function.
We provide an illustrative example of this generic behavior in  
Fig.~\ref{fig:probing} for a simple qubit case $D=2$
and $N=3$ phases. In principle, there can be more extrema but we did not find 
them for any reasonable cost function studied. We did however observe 
additional extrema when considering mutual information as the figure of merit 
as outlined in Sec.~\ref{sec:mutual}.




\section{Sub-covariant measurements}
\label{sec:subcovariant}
In order to put our results in a more rigorous mathematical framework, we provide here
a general theory of sub-covariant measurements, that help to remedy the
apparent contradictions one may arrive at when applying the theory of standard
covariant measurements to discrete parameter estimation problems, of which discrete phase estimation is a special case.
We would like to stress that this concept is completely general and may used in an arbitrary
discrete parameter estimation models beyond phase estimation. For this reason we
will formulate it in a general way, where an estimation model arises as a result
of choosing a  subgroup $H$ of a group $G$ that generates the original
continuous estimation problem. Note that we do not even insist that the group
$H$ is discrete here, as this will not be relevant for the general discussion,
hence the idea may applied even beyond the problem of discrete parameter
estimation.

Let $G$ be a group and $H$ be a subgroup of $G$. We associate elements of $H$
with possible true values of a quantum parameter encoded in a state and elements
of $G$ with
possible measurement outcomes which are also the possible estimated values of
the parameter.
By $e$ we denote the neutral element of $G$.
Let $U_g$ be a unitary representation of group $G$ on the Hilbert space of
interest.

\emph{Definition.} A POVM measurement $\{\Pi_{\tilde{g}}\}$ is
\emph{sub-covariant with respect
to subgroup $H \leq G $} if and only if
\begin{equation}
 \forall_{h\in H,\tilde{g}\in G}U_h\Pi_{\tilde{g}}U_h^\dagger=\Pi_{h\tilde{g}}.
\end{equation}

Let us observe that according to the above definition every measurement
sub-covariant with respect to subgroup $H$ is uniquely defined by a family of
operators $\{\Pi_\gamma\}_{\gamma\in G/H}$ by the following relation:
\begin{equation}
 \Pi_{\tilde{g}}=U_h\Pi_{\gamma}U_h^\dagger,
\end{equation}
where $\tilde{g}=h\gamma$.

\emph{Theorem.}
If the estimation problem is sub-covariant with respect to $H \leq G$, i.e.:
(i) $dh$---Haar measure on $H$, (ii) $\rho_h=U_h\rho_eU_h^\dagger$ for every
$h\in H$
 and (iii) $C(g''g',g''g)=C(g',g)$ for every $g,g',g''\in G$,
then we can choose the optimal measurement from measurements sub-covariant with
respect to $H\leq G$.

\emph{Proof:}
Let $\{\Pi^{\t{opt}}_{\tilde{g}}\}$ be the optimal measurement which minimizes
the average cost
\begin{equation}
\bar{C}(\Pi^{\t{opt}}_{\tilde{g}}) =
\int\limits_Hdh\int\limits_Gd\tilde{g}\Tr(\Pi^{\t{opt}}_{\tilde{g}}
\rho_h)C(h,\tilde{g}).
\end{equation}
 We define
\begin{equation}
\Pi^{\t{cov}}_{\tilde{g}}=\int\limits_Hdh'U_{h'}^\dagger\Pi^{\t{opt}}_{h'\tilde{
g }
}U_{h'}.
\end{equation}
Then $\Pi^{\t{cov}}_{\tilde{g}}$ is sub-covariant with respect to $H$ since:
\begin{multline}
U_h\Pi^{\t{cov}}_{\tilde{g}}U_h^\dagger=   U_h\int\limits_Hdh'U_{h'}^\dagger
\Pi^{\t{opt}}_{\tilde{g}}U_{h'}U_h^\dagger= \\
= \int\limits_Hdh'U_{hh'^{-1}}\Pi^{\t{opt}}_{
h'\tilde{g}}U_{h'h^{-1}}=
 \int\limits_Hdh'U_{h'}^\dagger\Pi^{\t{opt}}_{
h'h\tilde{g}}U_{h'}=\Pi^{\t{cov}}_{h\tilde{g}},
\end{multline}
where in the last equality we have we substituted $h' \rightarrow h'h$, making
use of the Haar measure property.
Moreover, measurement $\Pi^{\t{cov}}_{\tilde{g}}$ yields the same cost as
$\Pi^{\t{opt}}_{\tilde{g}}$, which can be seen as follows:
\begin{multline}
\bar{C}(\Pi^{\t{cov}}_{\tilde{g}})=
\int\limits_Hdh\int\limits_Gd\tilde{g}\Tr(\Pi^{\t{cov}}_{\tilde{g}}
\rho_h)C(h,\tilde{g})= \\
= \int\limits_Hdh\int\limits_Gd\tilde{g}\Tr(\int\limits_Hdh'U_{h'}^\dagger\Pi^{
\t{opt}
}_{h'\tilde{g}
}U_{h'}
U_h\rho_eU_h^\dagger)C(h,\tilde{g})= \\
= \int\limits_Hdh\int\limits_Gd\tilde{g}\int\limits_Hdh'\Tr(U_{hh'}^\dagger\Pi^{
\t{opt}
}_{h'\tilde{g}
}U_{hh'}\rho_e)C(h,\tilde{g})= \\
= \int\limits_Hdh\int\limits_Gd\tilde{g}\int\limits_Hdh'\Tr(U_{h}^\dagger\Pi^{
\t{opt}
}_{\tilde{g}
}U_{h}\rho_e)C(h'^{-1}h,h'^{-1}\tilde{g})=  \\
 = \int\limits_Hdh\int\limits_Gd\tilde{g}
\underbrace{\int\limits_Hdh'}_{=1}\Tr(\Pi^{
\t{opt}
}_{\tilde{g}
}\rho_h)C(h,\tilde{g})=\bar{C}(\Pi^{\t{opt}}_{\tilde{g}}),
\end{multline}
where in the third line we have substituted $ h\rightarrow
h'^{-1}h,\tilde{g}\rightarrow h'^{-1}\tilde{g}$.   This ends the proof
$\square$.

Unlike in the full-covariant estimation problem, the search for the optimal
measurement is not restricted here to identifying a single seed POVM $\Pi_0$,
but rather a whole set
$\Pi_\gamma$ where $\gamma \in G/H$. Still, this fact may significantly speed
up the numerical search for the optimal estimation strategy
as for sub-covariant measurements the average cost simplifies to:
\begin{equation}
\label{eq:costsubcov}
\bar{C} = \int\limits_{G/H} d\gamma \int\limits_H dh \Tr(U_h \Pi_\gamma
U_h^\dagger \rho) C(0, h \gamma).
\end{equation}

Let us now discuss the relation of the above abstract and general theory with 
the phase estimation problem, especially with the examples presented in 
Secs.~\ref{sec:discrimination} and~\ref{sec:general}. 
In the continuous phase estimation problem $H=U(1)$ and sub-covariant 
measurements are not useful since there is no relevant group $G$ of which $H$ 
would be a nontrivial subgroup. In the discrete phase estimation problem 
$H=U_N$, and 
$\int_H dh\rightarrow\sum_{n=0}^
{N-1}$, while in general $G=U(1)$. Hence the task is to find seed POVM set 
$\Pi_\gamma$, where $\gamma \in
[0,\theta) = U(1)/U_N$. However, is is not always necessary to consider the 
full $G=U(1)$ group as it might be the case that the cost is minimal already 
for another discrete group $G=U_{lN}$, $l\in\mathbb{Z}$. An example is 
the shifted-covariant strategy optimal for some values of $N$ in 
Secs.~\ref{sec:discrimination} and~\ref{sec:general}, which corresponds to 
$H=U_N$, 
$G=U_{2N}$ and the seed POVM set \{0, $D U_{\frac{\pi}{N}} \ket{e}\bra{e} 
U^\dagger_{\frac{\pi}{N}}$\}. 

Let us also note that in the examples of discrete phase estimation discussed in 
the preceding 
sections we have not identified situations where more than one nonzero seed 
POVM would be necessary, and the optimal performance could always be reached by 
shifted-covariant measurements, which are the simplest non-trivial example of 
sub-covariant strategies. We expect, however, that for more involved discrete 
estimation problems more general class of measurements might be required. 
Therefore, in the next Sec.~\ref{sec:algorithm} we 
provide a general algorithm to search for optimal sub-covariant strategies.

\section{Algorithm for identifying the optimal discrete phase estimation
strategy}
\label{sec:algorithm}
Here we show how to effectively use the concept of sub-covariant measurements
in order to numerically find the optimal discrete parameter estimation problems. 
For concreteness, we will focus on dicrete phase estimation, but the procedure 
can be directly generalized to any subgroup estimation problem using the 
framework described in Sec.~\ref{sec:subcovariant}.

First assume for the moment that the input state is fixed. We know that we can
look for the optimal
measurement/estimation strategy within the class of sub-covariant measurements.
Hence we need to find the optimal seed POVM set $\Pi_\gamma$, where $\gamma \in
[0,\theta) = U(1)/U_N$.
Clearly looking for continuous family of operators is not feasible numerically.
Still one can proceed in steps. Let us define sets containing an increasing
number
of seed measurement operators: $\mathcal{P}_1 = \{ \Pi_0\}$, $\mathcal{P}_2 = \{
\Pi_0, \Pi_{\frac{1}{2} \theta}\}$,
$\mathcal{P}_3 = \{ \Pi_0, \Pi_{\frac{1}{3} \theta}, \Pi_{\frac{2}{3} \theta}\}$
etc.  Considering a given number $S$ of seed measurements, we
minimize $\bar{C}$ over operators $\Pi_{s \theta/S}$, where $s \in \{0,\dots,
S-1\}$, with the following constraints:
$\Pi_{\frac{s}{S}\theta} \geq 0$, $\frac{1}{N}
\sum_{s=0}^{S-1} \sum_{n=0}^{N-1} U_{n \theta} \Pi_{\frac{s}{S} \theta}
U_{n\theta}^\dagger = \openone$.
This is a standard semi-definite program and can be solved efficiently using
e.g. the CVX package for Matlab \cite{cvx}. We proceed by increasing $S$ and
once we observe that further increase in $S$ does not reduce the cost we stop.

In order to find the optimal input state as well, one should adapt an iterative
approach, as proposed in \cite{Demkowicz2011, Macieszczak2014},  where the
search for the optimal measurements is interleaved with the search for the
optimal state. Applying the general form of Eq.~\eqref{eq:costsubcov} to our
case of discrete quantum phase estimation, we
obtain:
\begin{equation}
\bar{C} = \bra{\varphi_0} A \ket{\varphi_0}, \quad A =
\frac{1}{N}\sum_{s=0}^{S-1} \sum_{n=0}^{N-1} U_{n\theta} \Pi_{\frac{s}{S}
\theta} U_{n\theta}^\dagger
C(n \theta + \tfrac{s}{S} \theta).
\end{equation}
We see that  given a particular sub-covariant measurement, the search for the
state minimizing the mean cost is equivalent to the search of an eigenvector
corresponding to the minimal eigenvalue of
the matrix $A$. By iterating the above scheme of finding the measurement
optimal
for a particular input state, and the the state optimal for a particular
measurement one arrives at the fully optimal solution.

In particular, we have followed this procedure to confirm that the discrimination strategy described in
Sec.~\ref{sec:discrete}, based on switching between covariant and 
shifted-covariant measurements was indeed optimal.

\section{Mutual information as a figure of merit}
\label{sec:mutual}
In previous sections we have pursued an approach to quantum estimation/discrimination problems 
focused on minimization of Bayesian cost functions. This is a natural framework for determining optimal protocols 
in single-shot estimation/discrimination scenarios. One can, however, look at 
the discrete phase estimation setting as an 
example of a communication protocol, which is supposed to be repeated many times. In such situations a natural figure of merit is 
 the mutual information between the encoded phases and measurement results which 
quantifies how many bits of noiseless information 
  can be effectively transmitted per single use of the considered quantum channel.
  \begin{figure}[t!]
\includegraphics[width=\columnwidth]{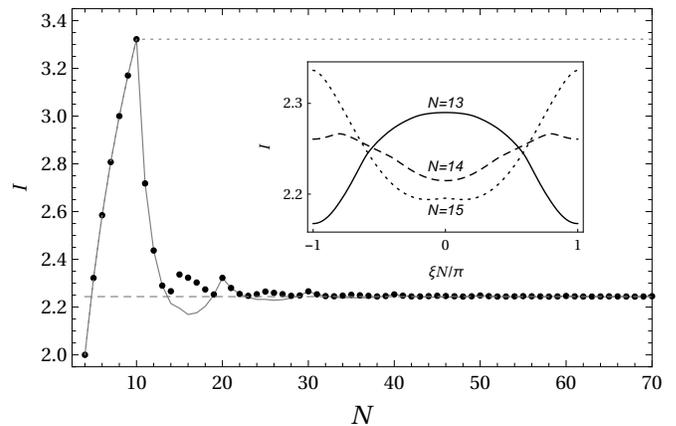}
\caption{Mutual information vs. number of phases depicted for a $D=10$ 
dimensional system. Black dots represent the shifted-covariant measurements 
optimized over the shift parameter $\xi$ for each $N$, while solid gray 
line corresponds to the covariant measurements ($\xi=0$). For $N<D$, the 
shifted-covariant and covariant strategies are equivalent and saturate the 
Holevo bound (with Holevo quantity shown as dotted gray line). For $N\geq D$ 
the covariant and shifted-covariant strategies cease to yield the same results; 
the mutual information is greater or 
equal the continuous limit (dashed gray) for any $N$ only for the optimal 
shifted-covariant strategy. The 
inset shows the mutual information vs. the shift parameter $\xi$; the maxima 
at $\xi=0$, $0< \xi < \pi/N,$ and $\xi=\pi/N$ are observed for the 
exemplary cases of 
$N=13$ (solid), $N=14$ (dashed) and $N=15$ (dotted), respectively.
}
\label{fig:mutual}
\end{figure}
Given equiprobable phase encoded input states $\ket{\varphi_n}=U_{\varphi_n} 
\ket{\varphi_0}$, $\varphi_n = n \theta$, $n=0, \dots N-1$, 
and POVM measurement $\left\{ \Pi_m \right\}$, the resulting joined probability distribution 
of symbol $n$ being sent and measurement outcome $m$ being registers reads:
\begin{equation}
p_{nm} = \frac{1}{N} \t{Tr}(\ket{\varphi_n}\bra{\varphi_n} \Pi_m),
\end{equation}
while the corresponding mutual information reads:
\begin{equation}
I = \sum_{mn} H\left(p_{nm}\right) - \sum_n H\left(\sum_m p_{nm}\right) - \sum_{m} H \left(\sum_{n} P_{nm}\right),
\end{equation}
where  $H(x) = -x \log_2 x$. The optimal protocol, from this point of view, is now the one that maximizes $I$ over the choice of input state
$\ket{\varphi_0}$ and measurements $\{\Pi_m\}$. 

 Unlike the  Bayesian cost, here $I$ is nonlinear both in measurement and the input state which makes the maximization of $I$ much more challenging. For covariant problems, there is no general theorem on the optimality of covariant measurements
 in this case, except for the situation when the states used are generated by 
the action of an irreducible representation of a group \cite{Davies1978, 
Sasaki1999}. In our case the $U(1)$ representation is clearly reducible and 
hence Davies theorem cannot be invoked. 
 Still, one can bound the maximal achievable mutual information using the famous Holevo bound \cite{Holevo1982}.  For our problem 
 this bound implies:
 \begin{equation}
 I \leq \chi = S\left( \frac{1}{N} \sum_{n=0}^{N-1} \ket{\varphi_n} \bra{\varphi_n}\right),
 \end{equation}
where $S(\rho) = - \t{Tr}\rho \log_2 \rho$ is the von Neumann entropy.
For $N \leq D$, maximum $\chi$ corresponds to situation where we have $N$ orthogonal states $\ket{\varphi_n}$ and $\chi = \log_2 N$. 
This bound can simply be achieved by using exactly the same strategy as 
described when minimizing Bayesian cost for $N \leq D$. 
Since $N$ orthogonal states are perfectly distinguishable by performing 
measurement in the basis containing the input states
we can obtain $I = \log_2 N$, saturating the Holevo bound and proving that this strategy is indeed optimal.
In general the Holevo bound can never be larger than $\log_2 D$ (maximum entropy is reached for $\rho= \openone/D$), 
hence for $N > D$ the bound has the form $I \leq \log_2 D$ and does not further increase with $N$. 
Even though this bound cannot be in general achieved, it has been shown \cite{Maccone2017} 
that in the limit of continuous phase estimation $N\rightarrow \infty$, and large dimensions $D \rightarrow \infty$ 
one can reach almost optimal performance $I \approx \log_2 D - 1.2199$. This performance can be achieved using
a strategy utilizing an equal superposition input states $\ket{\varphi_0} = \ket{e}= \frac{1}{\sqrt{D}}\sum_{k=0}^{D-1} \ket{k}$ and
standard covariant continuous POVM, $\Pi_{\tilde{\varphi}} = \frac{D}{2\pi} U_{\tilde{\varphi}}\ket{e}\bra{e} U_{\tilde{\varphi}}^\dagger$.

While studying discrete-to-continuous transition using mutual information figure 
of merit we will employ the above coding-decoding strategy. 
When going from the case $N=D$, where the mutual information can be made equal to the Holevo bound, to $N \rightarrow \infty$ (continuous regime) the accessible mutual information can only decrease, hence, we can sensibly expect that, since in the continuous limit the above described procedure performs almost optimally, it performs close to optimal also in the discrete case.
Similarly as in the Bayesian approach, however, we have observed that it is essential to consider shifted-covariant measurements to obtain optimal performance. Note that when dealing with mutual information, there is no issue with the set of estimated values, as we deal solely with probability distribution. Still, when considering the specific class of POVMs as described above, we can naturally define a shifted-covariant POVM as  $\Pi^\xi_{\tilde{\varphi}} = \frac{D}{2\pi} U_{\tilde{\varphi}+\xi}\ket{e}\bra{e} U_{\tilde{\varphi}+\xi}^\dagger$.
In Fig.~\ref{fig:mutual} we depict the results for the discrete-to-continuous transition for $D=10$. 
Interestingly, in order to reach the optimal mutual information we occasionally have encountered situations where the optimum was reached 
for $0< \xi < \pi/N$, unlike in Bayesian examples studied where either $\xi=0$ 
and $\xi=\pi/N$ shifts were optimal. This is due to much more more involved 
structure of the mutual information potential leading to more than two local 
extrema while changing the shift parameter $\xi$.  

We leave it as an open question, whether the above described strategy is optimal, and in particular 
whether more general measurements, that go beyond the class shifted-covariant measurements, could be useful in increasing the mutual information
which is known to be the case in some communication problems \cite{Shor2000}. 

\section{Conclusions}
\label{sec:conclusion}
In this paper, we have provided methods and results concerning the discrete phase estimation problem and its transition to the continuous limit.
We have studied the transition for different cost functions and showed that in
general one may need to go beyond the standard concept of covariant measurements
and use a more general form---sub-covariant measurements introduced in this
work.

We believe that these results may find their applicability in quantum
communication  theory where phase shift keying protocols involve encoding a
discrete set of phases on the transmitted states.
We also see this work as a first step towards better understanding a general
problem of digital-to-analog transition in decoding and encoding information
using quantum states, e.g. in quantum reading protocols \cite{Pirandola2011}.

\section*{Acknowledgments}
We thank Michael J. W. Hall for pointing out the
relationship between Sec.~\ref{sec:discrimination} and
Discrete Prolate Spheroidal Sequences. This work was supported by the  Polish
Ministry  of  Science  and  Higher
Education  Iuventus  Plus  program  for  years  2015-2017 No.   0088/IP3/2015/73.

\end{document}